\documentclass[twocolumn, traditabstract]{aa} 
\usepackage{graphicx}
\usepackage{amsmath} 
\usepackage{subeqnarray}
\usepackage{layouts}
\usepackage{color}
    \definecolor{Blue}{rgb}{0.0,0.0,1.0}
    \definecolor{Red}{rgb}{1.0,0.0,0.0}
    \definecolor{Green}{rgb}{0.0,0.0,1.0}
\begin{document}
\title{Eddington Capture Sphere around luminous stars}
\author{      Adam Stahl\inst{1} %
  \and Maciek Wielgus\inst{2} %
  \and \\ Marek Abramowicz\inst{1,3,4} %
  \and W{\l}odek Klu{\'z}niak\inst{3}
  \and Wenfei Yu\inst{5} }
\institute{    Physics Department, Gothenburg University, SE-412-96
  G{\"o}teborg, Sweden \\ \email{gusstaad@student.gu.se}
  \\ \email{marek.abramowicz@physics.gu.se}
  \and       Institute of Micromechanics and Photonics, ul. {\'s}w.
  A. Boboli 8, PL-02-525, Warszawa, Poland
  \\ \email{maciek.wielgus@gmail.com}
  \and          Copernicus Astronomical Center, ul. Bartycka 18,
  PL-00-716 Warszawa, Poland \\ \email{wlodek@camk.edu.pl}
  \and           Institute of Physics, Faculty of Philosophy and
  Science, Silesian University in Opava, Bezru{\v c}ovo n{\'a}m. 13,
  CZ-746-01 Opava, Czech Republic
  \and           Shanghai Astronomical Observatory 80 Nandan Road CN
  Shanghai 200030, China \\ \email{wenfei@shao.ac.cn} }
\date{Received ????; accepted ????}
\abstract {Test particles infalling from infinity onto 
a compact spherical star with a mildly 
super-Eddington luminosity at its
surface are typically trapped on the ``Eddington 
Capture Sphere" and do not reach the 
surface of the star. The presence of
a sphere on which radiation pressure
balances gravity for static particles
was first discovered some twenty five years
ago.
Subsequently, it was shown to be a capture
sphere for particles in radial motion,
and more recently 
also for particles in non-radial 
motion, in which the Poynting-Robertson 
radiation drag efficiently removes the 
orbital angular momentum of the particles,
reducing it to zero. Here we develop this
idea further, showing that ``levitation" 
on the Eddington sphere (above the stellar surface)
is a state of stable equilibrium, and discuss
its implications for Hoyle-Lyttleton accretion onto
a luminous star. When the Eddington sphere is present,
the cross-section of a compact star for actual accretion
is typically less than the geometrical cross-section
$\pi R^2$, direct infall onto the stellar surface
only being possible for relativistic particles,
with the required minimum 
particle velocity at infinity typically $\sim1/2$ the speed
of light.  We further show that particles
on typical trajectories in the vicinity 
of the stellar surface will also be trapped 
on the Eddington Capture Sphere.}
\maketitle

\section{Introduction}
In Newtonian theory both gravity and radiative flux {\it in vacuo}
diminish like $1/r^2$ with the distance from the center of the
star. The luminosity in units of the Eddington luminosity,
$L(r)/L_{\rm Edd}$, at {\it any} distance from the star is equal to
its value at the surface, $L(R)/L_{\rm Edd}$, and is therefore 
{\it  everywhere} either sub-Eddington, Eddington, or
super-Eddington. However, in Einstein's general relativity the
radiative force diminishes more strongly with distance than the
gravitational force, the redshifted luminosity $L(r)$ decreasing as
$1/(1-2R_G/r)$, and a static balance of forces is achieved at that
radius $r$ at which $L(r)=L_{\rm Edd}(1-~2R_G/r)^{-1/2}$. For this
reason, radiation may be super-Eddington close to the star, but
sub-Eddington further away, reaching the Eddington value at the
``Eddington Capture Sphere'', whose radius is (\cite{Phinney})
\begin{equation}
\label{Eddington-radius}
R_{\rm Edd} =  \frac{2 R_G}{1 - \left(1 - \dfrac{2 R_G}{R} \right)^2
\left(\dfrac{L(R)}{L_{\rm Edd}} \right)^2}\ .
\end{equation}
At this radius the gravitational and radiative forces balance (for a~static,
optically thin hydrogen shell). The formula is in Schwarzschild
coordinates, with the gravitational radius $R_G$ defined in
Eq.~(\ref{dimensionless-scaling}).
\cite{Abramowicz} (hereafter AEL) have rigorously shown by analytic
calculations that in the combined gravitational and isotropic radiation fields of a~spherical, non-rotating, compact star,  radially moving test particles (with proton mass, and Thomson cross-section for photon momentum absorption) are captured on this sphere and ``levitate,'' i.e., remain at rest.

\cite{Oh2011} have shown that when the  luminosity at the surface of
the star is  mildly super-Eddington, particles with non-zero orbital
angular momentum can  also be captured at $R_{\rm Edd}$ (see also
\cite{Bini}). This is because the  Poynting-Robertson radiation drag
acts as an effective torque,  and for a wide class of trajectories
reduces the orbital  angular momentum to zero. 

In this paper 
we further discuss the phenomenon of the Eddington Capture Sphere
by considering two particular
cases, one corresponding to initially 
quasi-circular motion close to the star, 
and the second being rather similar
to the classic \cite{Hoyle} 
accretion model. We also
consider the question of stability and speculate 
about possible astrophysical
manifestations of the Eddington
Capture Sphere.

All numerical results presented here were obtained with the
Dormand-Prince method, which is a fourth-order accuracy,
adaptive step-size Runge-Kutta type integration method.

\section{Equations of motion}
We carry out all calculations in the Schwarzschild metric using
standard Schwarzschild spherical coordinates,
\begin{eqnarray}
\label{Schwarzschild-coordinates}
t &=& {\rm time}, ~~~r= {\rm radius}, \nonumber
\\
\theta &=& {\rm polar~angle}, ~~~\phi= {\rm azimuthal~angle},
\end{eqnarray}
and extend the treatment of AEL to non-radial motion. Spherical symmetry of the problem
assures that the trajectory
of a test particle
is confined to a single plane for any set of initial conditions.
Without loss of generality, we assume that this is the equatorial plane 
$\theta = \pi/2$ in
the Schwarzschild coordinates. Note that the stress-energy tensor of radiation
$T^{(i)(k)}$ calculated in the stationary observer tetrad by AEL does not depend on the
particle motion. Therefore, we may use here the AEL formulae for $T^{(i)(k)}$ with no
change. We use dimensionless radius $x$, and dimensionless proper time $\tau$ defined by the scaling of the line element $ds$, as well as the dimensionless
radius of the star, $X$,
\begin{equation}
d\tau = \frac{ds}{R_G}, \ x = \frac{r}{R_G},\ X = \frac{R}{R_G},
       ~~{\rm with}~ R_G = \frac{GM}{c^2},
\label{dimensionless-scaling}
\end{equation}
and introduce $B = 1 - 2/x$.


Assuming that the test particle is capturing momentum at a rate
proportional to the radiative flux in its rest frame, thus suffering
a rest-frame force of $(\sigma/c)\times$ flux, the particle trajectory
is described by two coupled, second order differential equations:
\begin{eqnarray}
\frac{d^2 x}{d \tau^2}
\!&=&\! \frac{k}{ \pi I(R) X^2}\ \left(BT^{(r)(t)}v^t -
\left[ T^{(r)(r)}+ \varepsilon \right]\frac{d x}{d \tau} \right)+
\nonumber \\
&&+ \left(x-3\right)\left( \frac{d\phi}{d \tau} \right)^2 - \frac{1}{x^2},
\label{system-1}\\
\frac{d^2 \phi}{d \tau^2} \!&=&\!
 - \frac{d \phi}{d \tau} \left(\frac{k}{\pi I(R) X^2}
 \left[T^{(\phi)(\phi)} + \varepsilon \right]
 + \frac{2}{x} \frac{dx}{d\tau}\right),
\label{system-2}
\end{eqnarray}
with the time component of the particle four velocity
\begin{equation}
v^t = B^{-\frac{1}{2}}\left[1 + B^{-1} \left(\frac{dx}{d\tau}\right)^2 + x^2 \left(
\frac{d \phi}{d \tau} \right)^2 \right]^{\frac{1}{2}},
\end{equation} and
\begin{eqnarray} \nonumber
\varepsilon \!&=&\!
B T^{(t)(t)}(v^t)^2 + B^{-1}T^{(r)(r)}
\left( \frac{dx}{d\tau}\right)^2 \\
&& + x^2 T^{(\phi)(\phi)} \left(\frac{d \phi}{d \tau} \right)^2 -
2T^{(r)(t)}v^t \frac{dx}{d\tau}.
\end{eqnarray}
The parameter $k$ is the surface luminosity in Eddington units,
\begin{equation}
k = \frac{L(R)}{L_{\rm Edd}} = 4\pi^2 R^2 \frac{I(R)}{L_{\rm Edd}},
\end{equation}
where $L_{\rm Edd} = 4 \pi G M m_p c/ \sigma$ .

\section{The capture sphere}
The system of Eqs. (\ref{system-1}), (\ref{system-2}) allows a static
solution \mbox{$x(\tau)=x_{\rm Edd}$} for
\begin{equation}
k = \left(1 - \frac{2}{x_{\rm Edd}} \right)^{\frac{1}{2}}
\left( 1 - \frac{2}{X}\right)^{-1},
\label{xn}
\end{equation}
where $x_{\rm Edd} = R_{\rm Edd}/R_G$, cf. Eq.~(\ref{Eddington-radius}). In the case of purely radial motion it was shown (AEL) that a critical point of nodal type is present at $x =x_{\rm Edd}$, resulting in a~stable equlibrium state of captured particles.  A~corresponding set of equilibrium points can be recognized in the system of Eqs. (\ref{system-1}), (\ref{system-2}). The set consists of points with $x =x_{\rm Edd}$, and any value of $\phi$ and $\theta$. 
(From Eq.~[\ref{system-2}] it follows that if $d \phi/dt = 0$,
 then $ d^2 \phi/dt^2 = 0$,
and Eq.~[\ref{system-1}] reduces to the radial equation of AEL.)
We refer to this set as the {\it Eddington Capture Sphere}. The radius of the sphere changes from $x_{ \rm Edd} = X$ to $x_{\rm Edd} = \infty$ as $k$ increases from $\left(1 - 2/X \right)^{-1/2}$ to $\left(1 - 2/X \right)^{-1}$.

\section{Stability of equilibrium}

\begin{figure}
\centering
\includegraphics[width=0.53\textwidth]{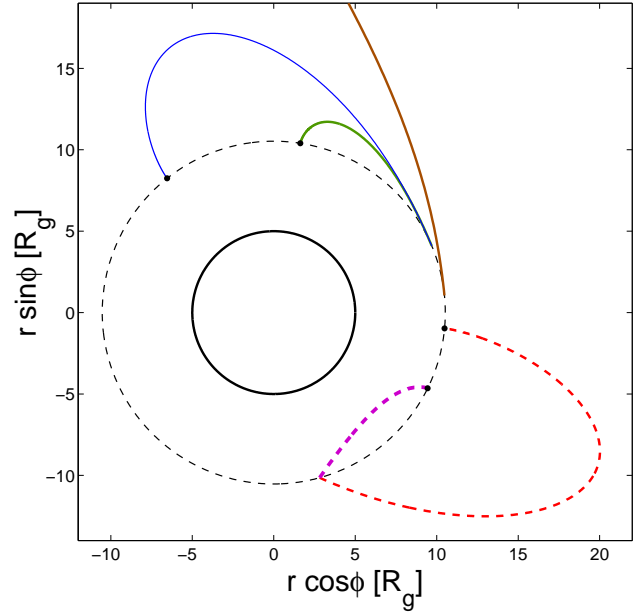}
\caption{Equilibrium on the Eddington Capture Sphere.
  Trajectories of particles returning to the Eddington sphere
  ($X=5$, $k=1.5$) are shown for four initial velocity perturbations: 
  $v^\phi \equiv x d\phi/d \tau = 0.12, v^r \equiv dx/d \tau = 0$ 
  (thick green continuous
  line),  $v^\phi = 0.20, v^r = 0$ (thin blue continuous line),
  $v^\phi = 0.12, v^r = 0.10$ (thick red dashed line) and  
  $v^\phi = 0.12, v^r = -0.10$ (extra thick violet dashed line).
  Also shown is a tangential escape trajectory (in thick brown) with initial
  $v^\phi = 0.32, v^r = 0$. }
\label{Fig:stability}
\end{figure}

Radial velocity perturbations were already discussed in AEL. In the
case of general velocity perturbations,  we observe stability in
numerical experiments: if  the azimuthal component of the velocity of
a particle located on the capture sphere is slightly (or not so
slightly) perturbed from zero value, the particle changes its location
but returns to the stationary state somewhere else on the sphere.  In
this sense, motionless ``levitation'' at a~distance $R_{\rm Edd}$ from
the star is a~stable state. Such behavior is presented in
Fig.~\ref{Fig:stability} for  $k = 1.50$, $X = 5.00$ (thick continuous
black line), corresponding to  $x_{\rm Edd} = 10.5$ (thin dashed black
line). We only show a small subset of the trajectories that we have
computed, and for clarity we only show orbits with large initial
velocity perturbations $\sim0.1c$.  For smaller perturbations the
result is the same.

In fact, one can compute the minimum escape velocities
from the Eddington Capture Sphere, these correspond
to radial trajectories and are $v^r_+=0.23$ for outward motion
(escape to infinity) and $v^r_-=0.51$ for inward motion (capture by the star),
for the parameters considered.
The minimum escape speed for trajectories tangent to the Eddington sphere
(the second cosmic speed) is $v^\phi=0.32$, the corresponding trajectory
is also exhibited (in brown) in Fig.~\ref{Fig:stability}.
One of the reasons for these
unexpectedly large values of the escape velocities
is the drag experienced by the particle even
in purely radial motion.

\section{Capture from circular orbits in the stellar vicinity}
In the absence of radiation,
a particle with appropriate angular momentum will follow a circular orbit around the star, but in a strong radiation field the particle will spiral inwards,
since the radiation drag continuously removes angular momentum. 

\begin{figure}
\centering
\includegraphics[width=0.48\textwidth]{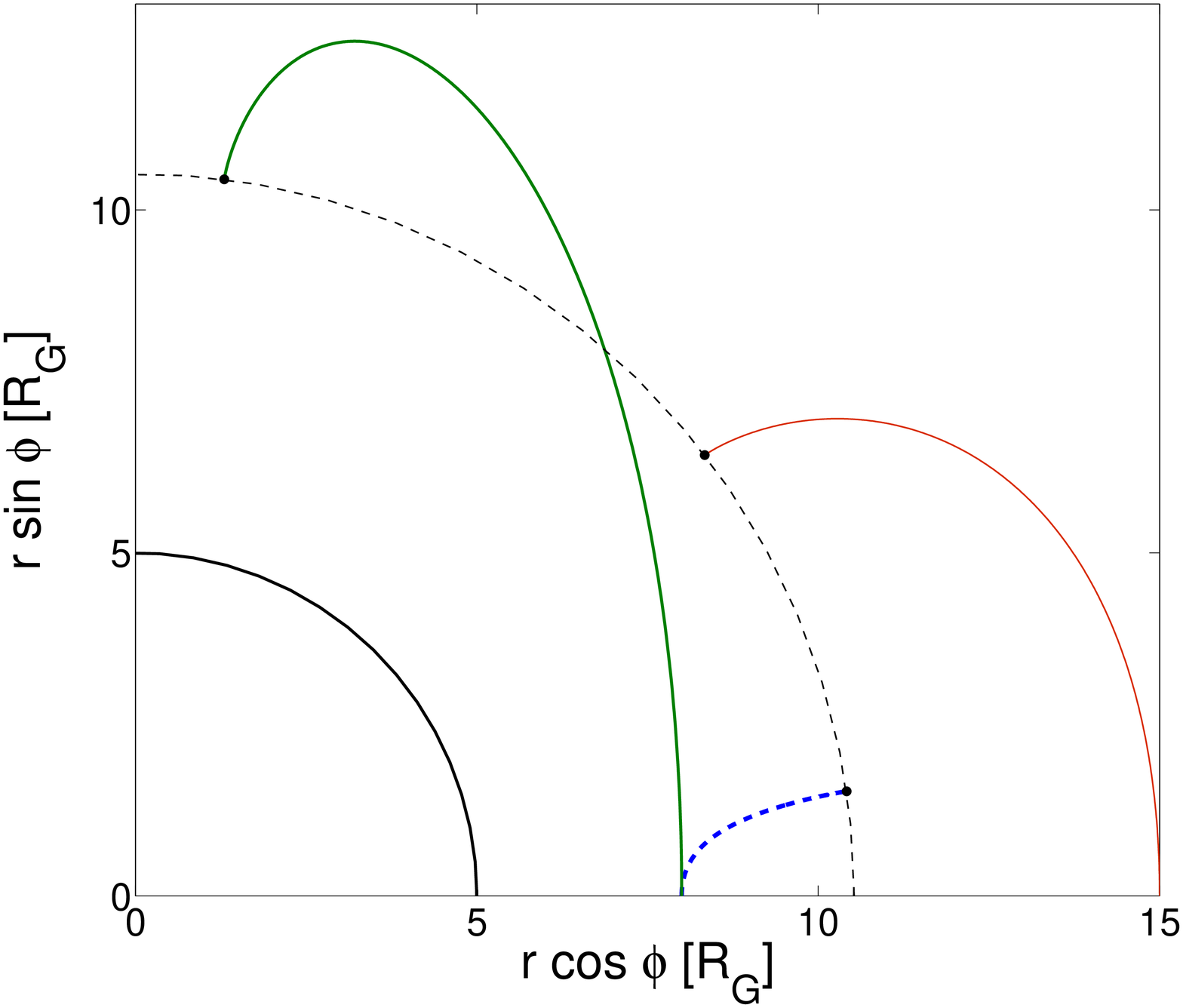}
\caption{Different trajectories for purely tangential initial velocity.}
\label{Fig:circular}
\end{figure}

\begin{figure}
\centering
\includegraphics[width=0.48\textwidth]{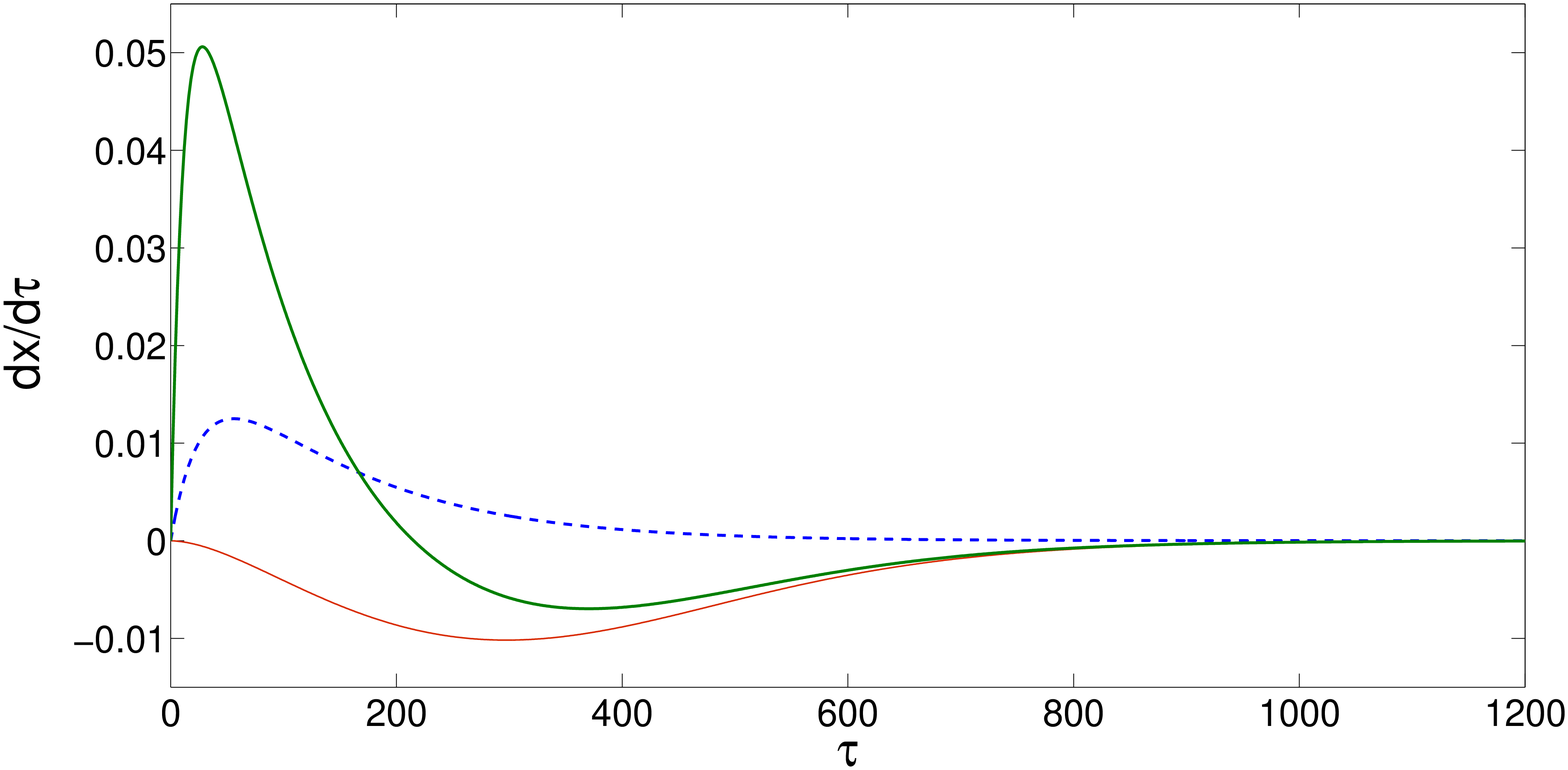} 
\includegraphics[width=0.48\textwidth]{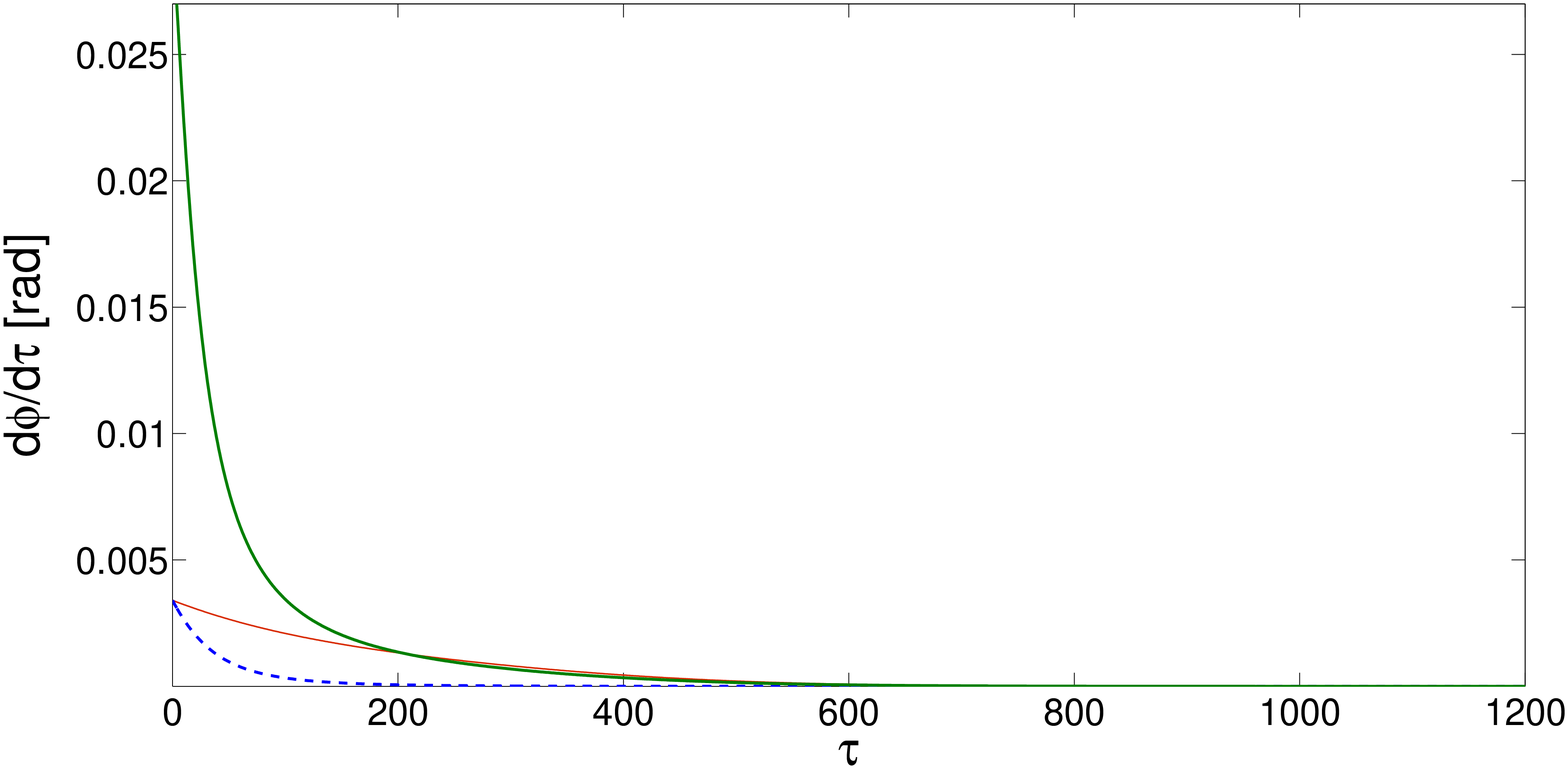}
\caption{Evolution of radial and angular velocity for trajectories in Fig. \ref{Fig:circular}.}
\label{Fig:circularEv}
\end{figure}

From 
Eq.~(\ref{system-1}), the azimuthal speed, $v_{\text{circ}}$,
corresponding to a trajectory
\emph{initially} tangent to a circle, may be derived.
Figure \ref{Fig:circular} shows some (non-escaping) trajectories ending on the
Eddington sphere at $x_{\rm Edd} = 10.5$ ($X=5.00$, $k=1.50$), 
 for particles with an initial zero 
value of radial component velocity, $dx/d\tau\,(0)=0$. Corresponding radial and angular velocity plots are presented in Fig. \ref{Fig:circularEv}. The trajectory shown with a thin continuous red line starts at 
$x_0 = 15.0$ with $v^\phi = v_{\text{circ}} = 0.00344$. 
The thick dashed blue one and the thick continuous green one start at 
$x_0 = 8.00$ with $v^\phi=0.00344$ and
 $v^\phi=0.0300$, respectively. They are all (eventually) attracted to the Eddington capture sphere, and in the end the approach is almost radial due to the effective removal of all angular momentum by the radiation field.

\section{Hoyle-Lyttleton accretion}

\begin{figure}
\centering
\includegraphics[width=0.48\textwidth]{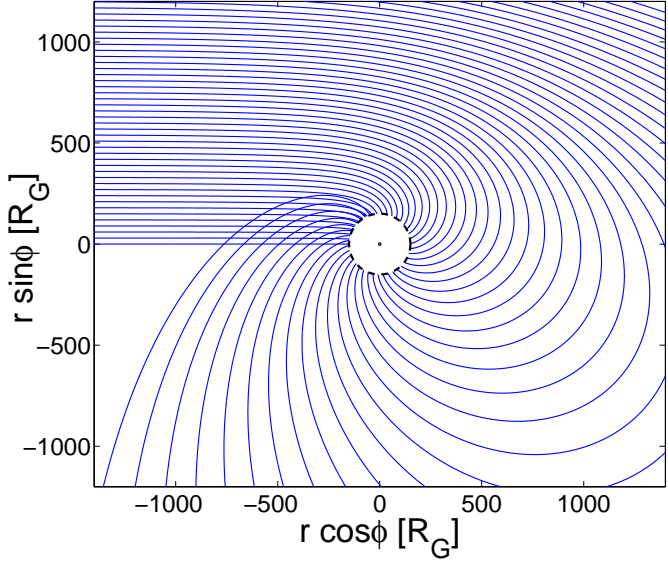}
\caption{The Hoyle-Lyttleton accretion problem for X=6.00, and k=1.49.
The trajectories of particles for $v_{\infty} = 5 \cdot 10^{-3} c$,
are shown. None of the test particles is accreted by the star,
they are either trapped on the Eddington sphere or escape to infinity.}
\label{Fig:HL}
\end{figure}

\begin{figure}
\centering
\includegraphics[width=0.48\textwidth]{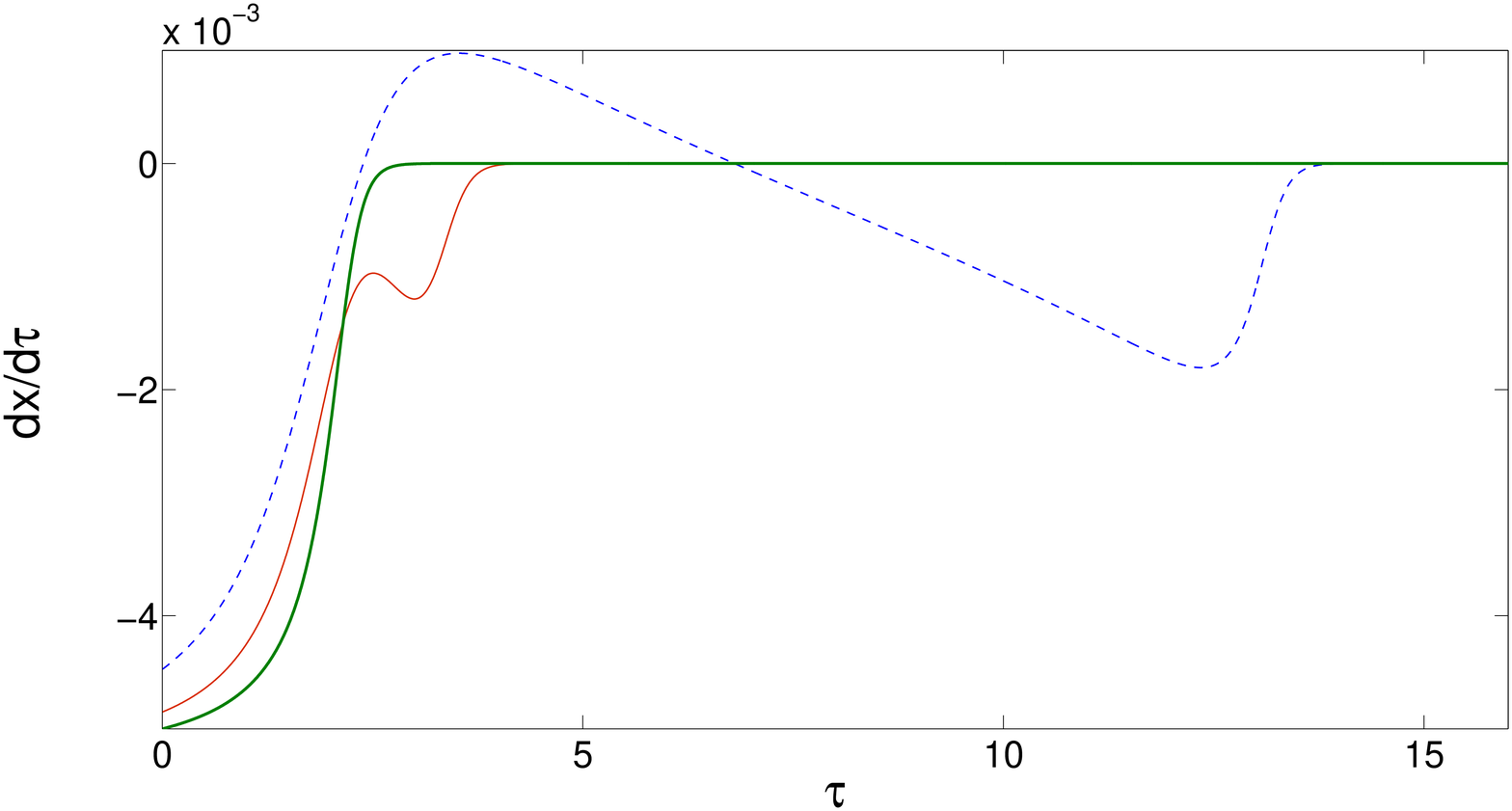}
\includegraphics[width=0.48\textwidth]{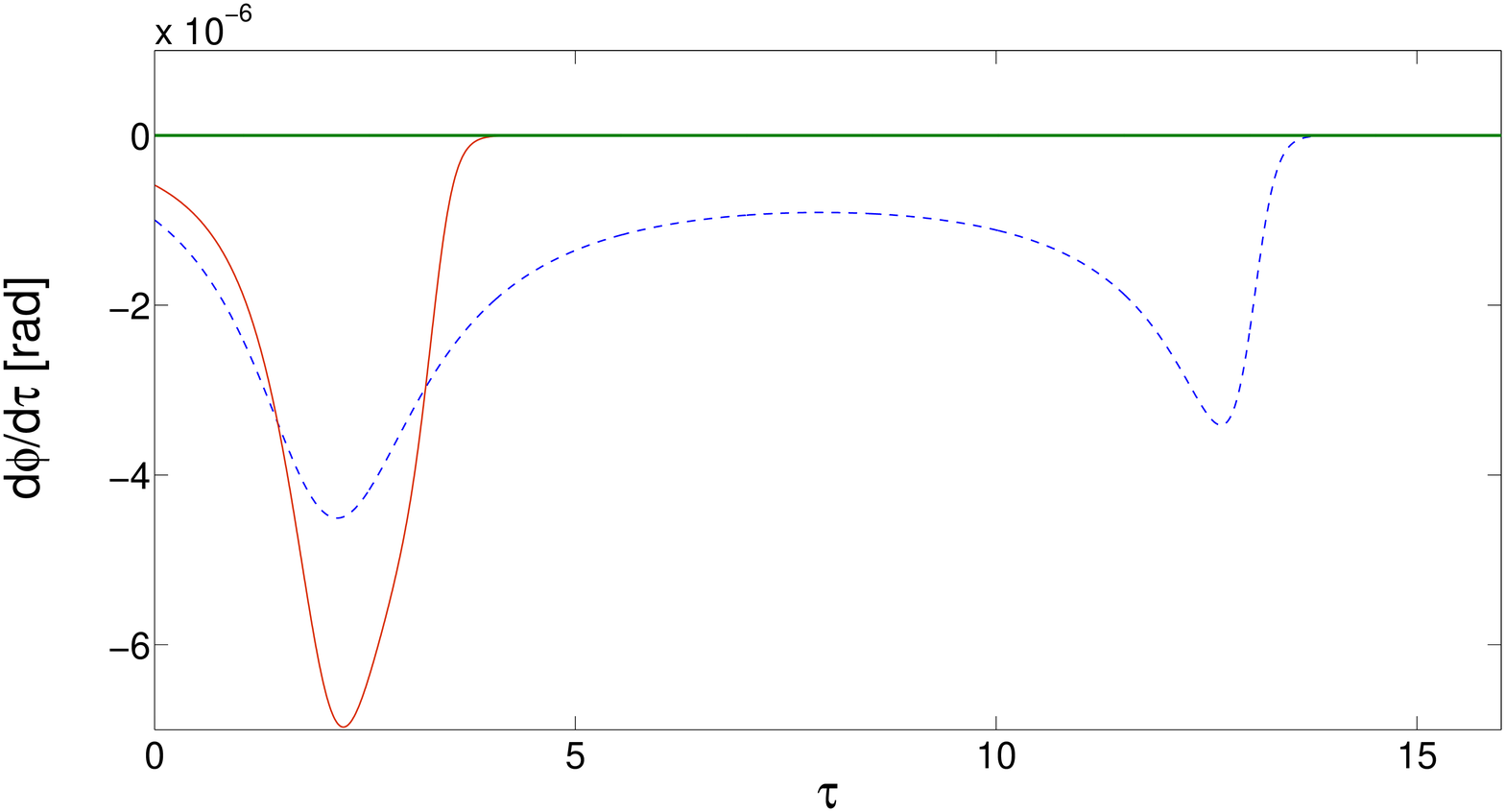}
\caption{Evolution of radial and angular velocity for three trajectories from
Fig.~\ref{Fig:HL} with impact parameter equal to 0 (thick continuous green), 
500 (thin continuous red) and 1000 (dashed blue line).}
\label{Fig:HLEv}
\end{figure}
We consider a situation resembling the model of a~star moving through a~uniform gas cloud, introduced by \cite{Hoyle}. In this Section we take the star
to have radius $X=6$.

In Fig.~\ref{Fig:HL}, numerically calculated trajectories of 
incoming particles are shown. The parameters
used were $k = 1.49$ (which implies \mbox{$x_{\rm Edd} = 151$}), and 
the initial velocity at infinity \mbox{$v_{\infty} = 5.00 \cdot 10^{-3}c$} 
(parallel to the $x$ axis) for all particles. The initial
value of $r \cos \phi$ was taken to be $-5000R_G$ and the impact
parameters (initial $ r \sin \phi$) of presented trajectories were
taken from the set $\{0, 30 R_G, 60 R_G, ... , 1500 R_G \}$. 
Radial and angular components of velocity are presented in 
Fig.~\ref{Fig:HLEv} for three selected particles, as a function of $\tau$.
The reduction of velocity along the trajectory
is attributed to the influence of radiation, bringing the particles to a halt
at $x = x_{\rm Edd}$. Close to the equilibrium sphere, motion is almost radial and
trajectories become perpendicular to the surface of the sphere. 

\begin{figure}
\centering
\includegraphics[width=0.48\textwidth]{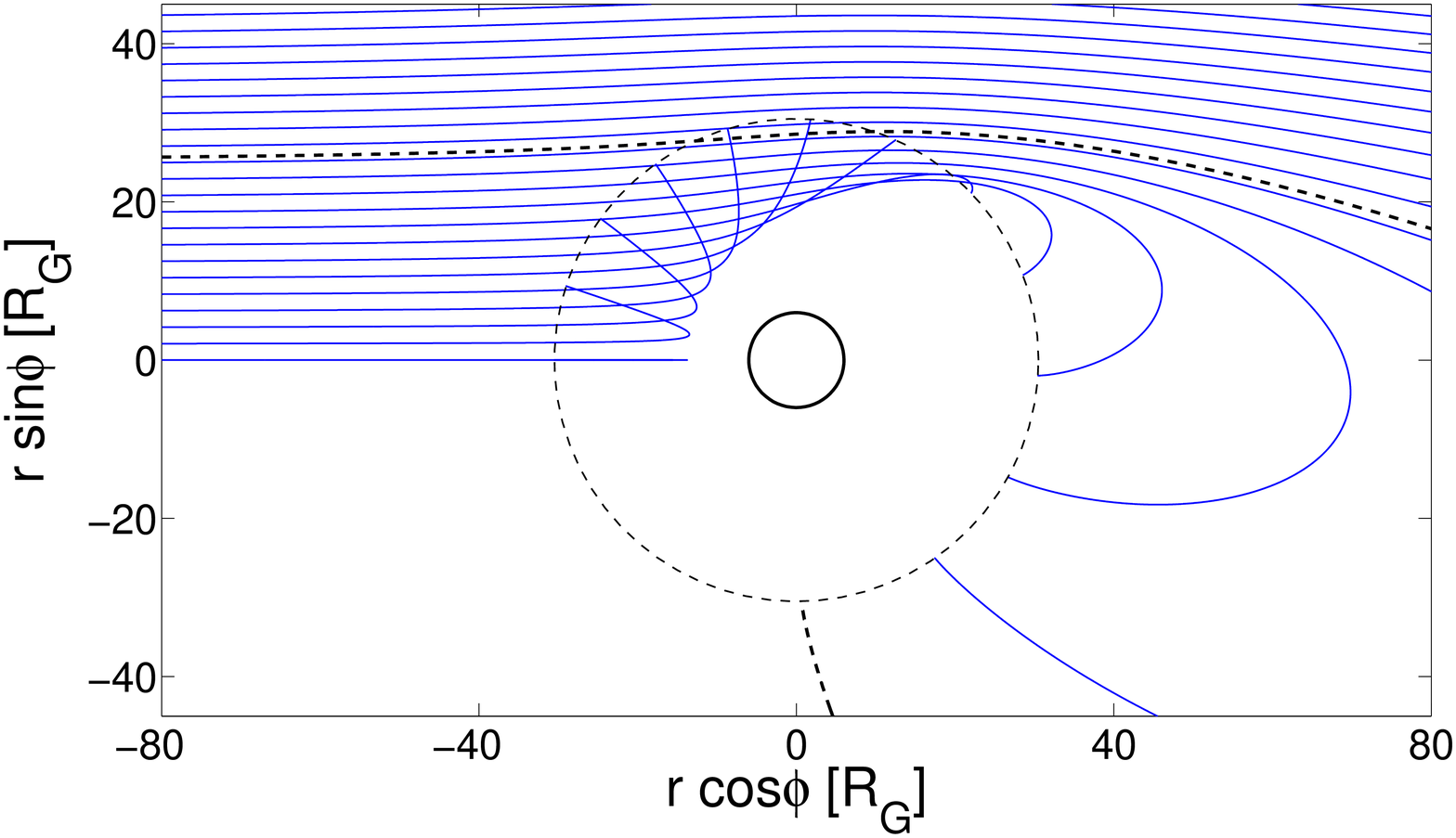}
\includegraphics[width=0.48\textwidth]{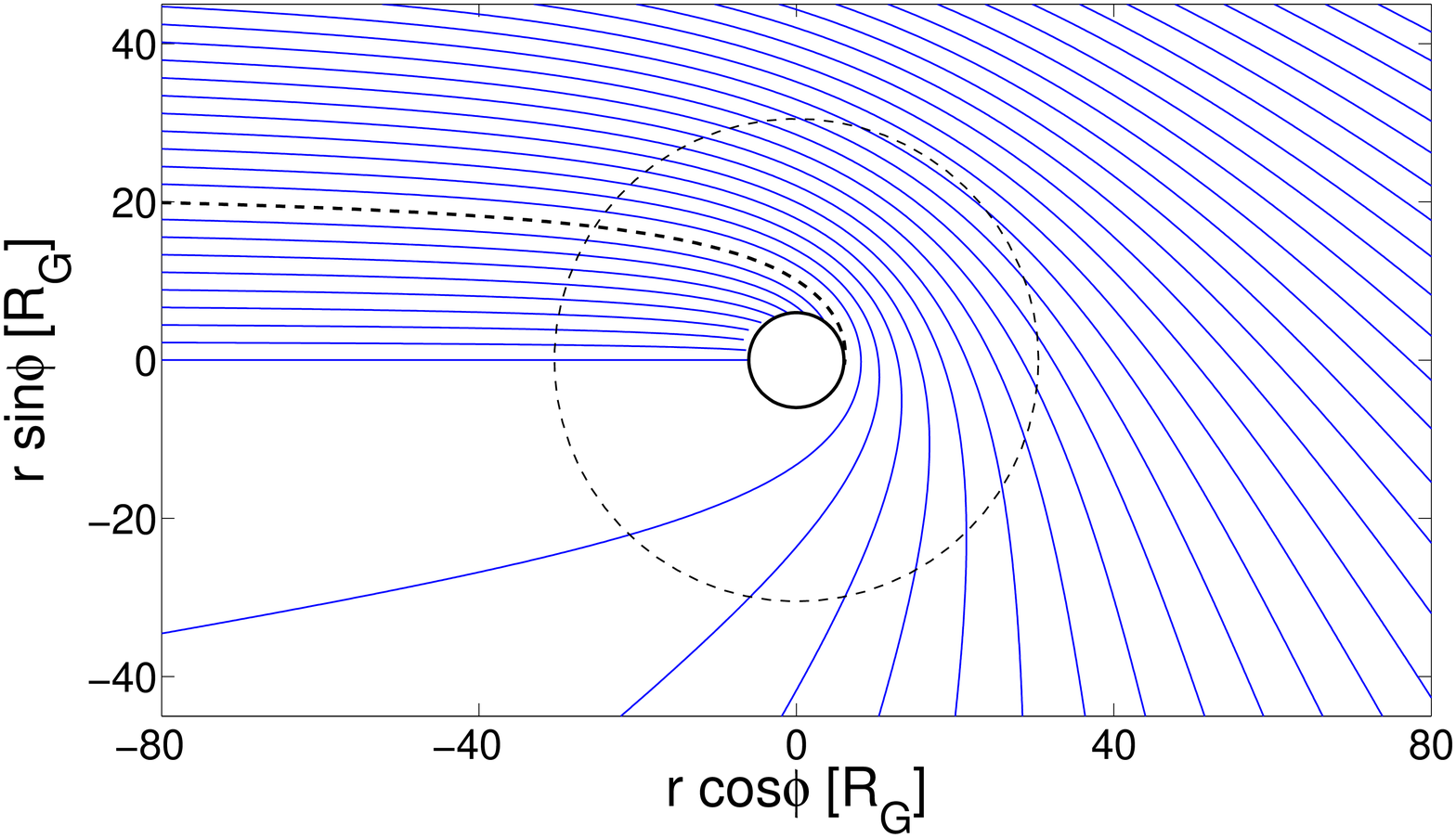}
\caption{Trajectories of particles for $v_{\infty} = 0.2 c$ with (top, k = 1.45) and without (bottom, k = 0) radiation influence. The stellar surface is indicated by the inner circle (thick, continuous line), and the Eddington sphere
by the dashed circle.}
\label{Fig:Penetrate}
\end{figure}

For a certain range of  kinetic energies, the incoming particles can
penetrate the capture sphere without  reaching the stellar
surface. Inside the sphere, radiation dominates over gravity and the
particles will be pushed out until they reach the surface of the
capture sphere again, and  come to rest there. This is illustrated in
Fig.~\ref{Fig:Penetrate}, with the parameters chosen to be
$v_{\infty} = 0.200c$, $X = 6.00$, $k = 1.45$, yielding
  $x_{\rm Edd} = 30.5$. The initial value of 
$r \cos \phi$ was equal to $-2000R_G$ 
and the impact parameters of presented
trajectories were taken from the set 
$\{0, 2 R_G, 4 R_G, ... , 50 R_G\}$.

The largest initial impact parameter of the captured particles was approximately equal to 1400 and 25.0 in the two
presented cases (Fig.~\ref{Fig:HL} and upper panel of Fig.~\ref{Fig:Penetrate}), while the critical impact parameters of Hoyle-Lyttleton accretion in the absence of radiation would have been equal to 849 and 22.4, respectively. Critical trajectories separating the trajectories of escaping particles from the captured ones (impact parameters equal to 25.0 and 22.4) are drawn with thick dashed lines in the upper and lower panel, respectively, of
 Fig.~\ref{Fig:Penetrate}. 

As noted by \cite{Oh}, in general, radiation drag is responsible for
capturing particles from a larger area than in the case of accretion
with no radiation. However, for the trajectories considered so far, no
particle falls onto the surface of the star. For particles actually
reaching the stellar surface, the converse statement is true, they
come from a smaller cross-sectional area than in the case of accretion
with no radiation, and in the cases considered here, smaller even than
the cross-sectional area of the star $\pi R^2$.

\subsection{Direct capture onto the star}
\begin{figure}
\centering
\includegraphics[width=0.48\textwidth]{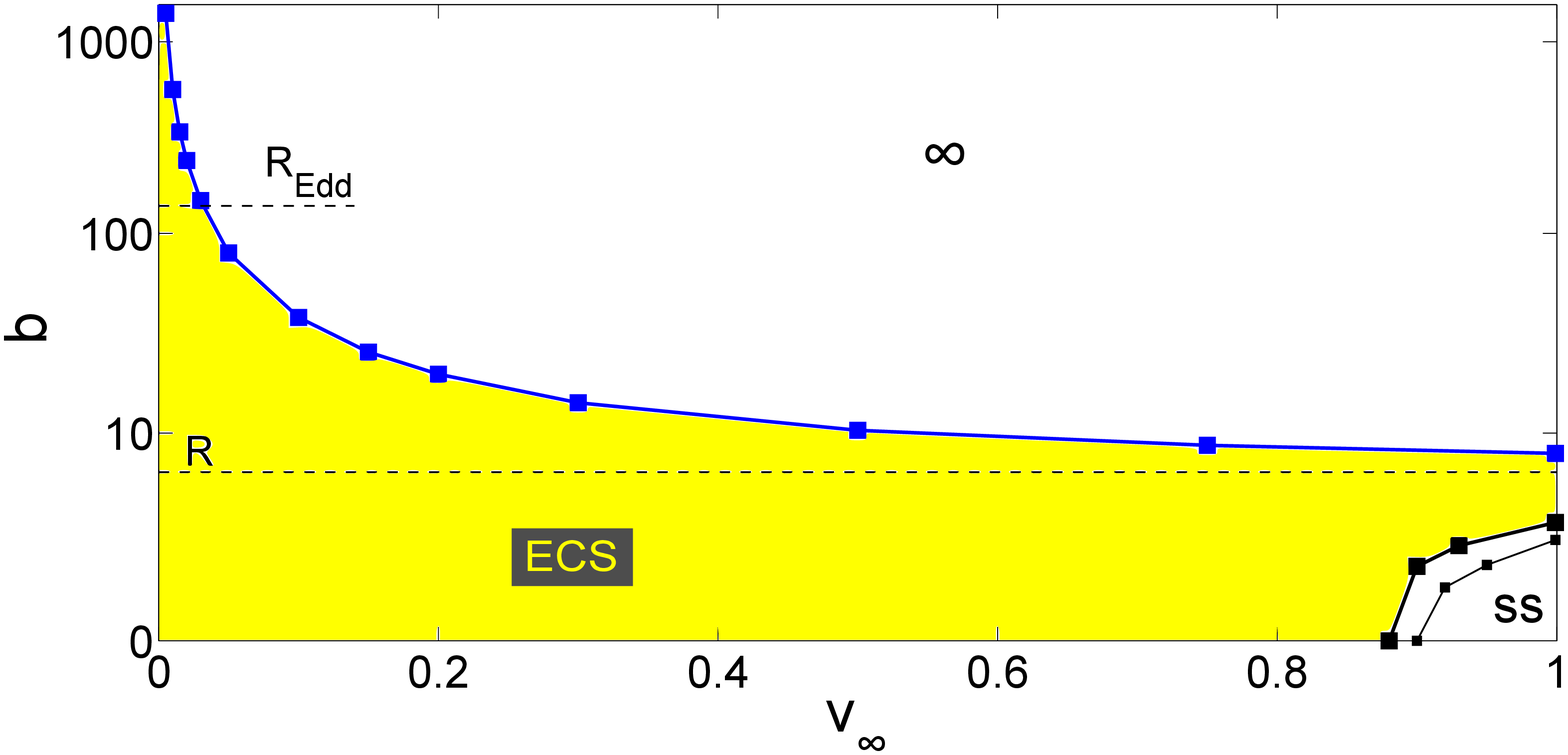}
\includegraphics[width=0.48\textwidth]{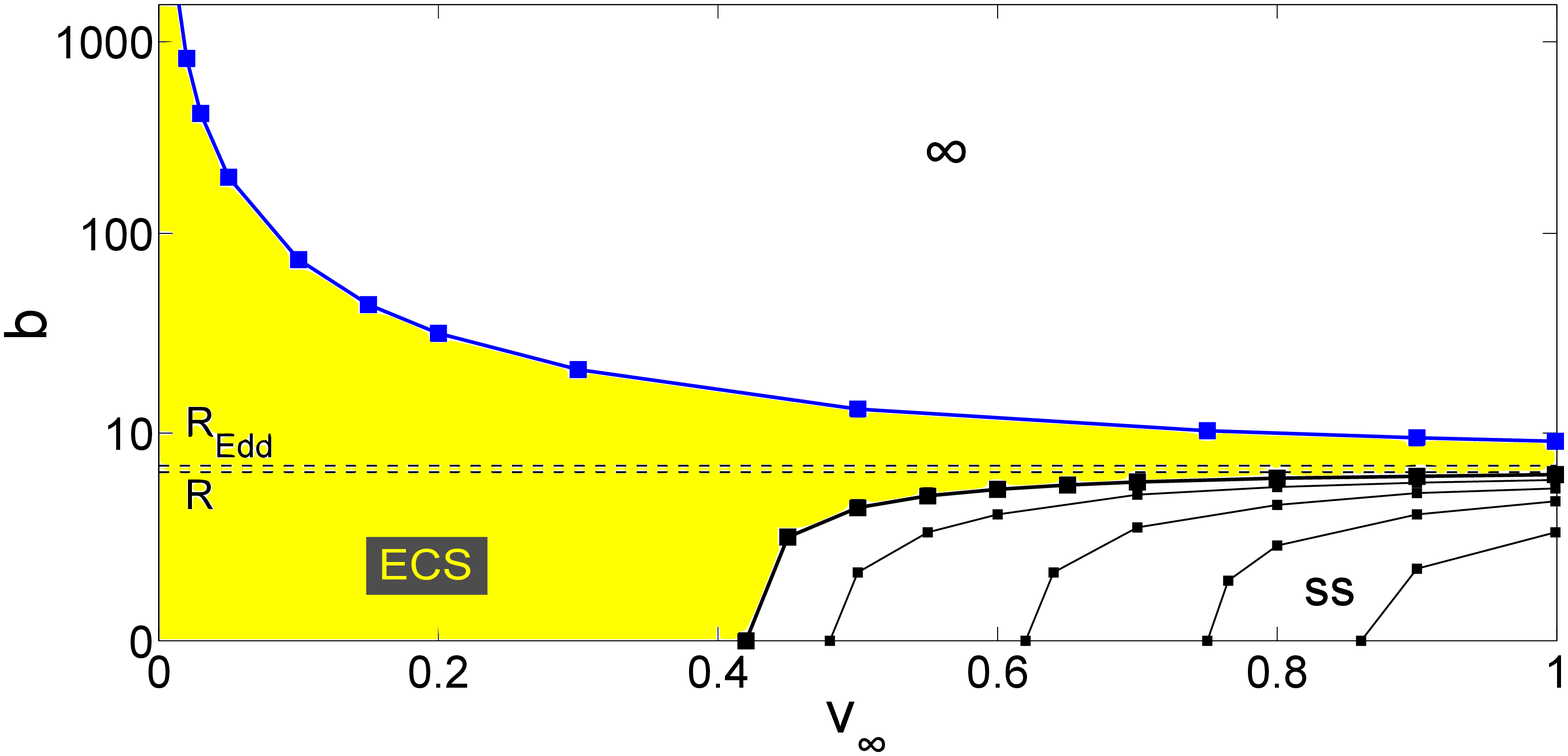}
\caption{Regions in the parameter space corresponding to escaping particles
(region denoted by ``$\infty$'' above the blue curve),
particles ending up on the Eddington Capture Sphere (the shaded
region between the thick
black curve and the blue curve, denoted by ``ECS''),
and  particles impacting the stellar surface 
(region below the thick black curve, denoted by ``ss''),
for two values of luminosity, k = 1.49 (top), and k = 1.25 (bottom).
The radius of the star is X=6 for all plots in this Section.
The impact parameter, $b$, is in units of $R_G$, and $v_\infty$
in units of $c$.
Particles with initial parameters on the thick black curve
settle on the surface of the star with zero velocity.
The thin black curves correspond to particles impacting with 
(from left to right): $v^r(R)=0.05,0.10,0.15, 0.20$.}
\label{Fig:ss}
\end{figure}

\begin{figure}
\centering
\includegraphics[width=0.48\textwidth]{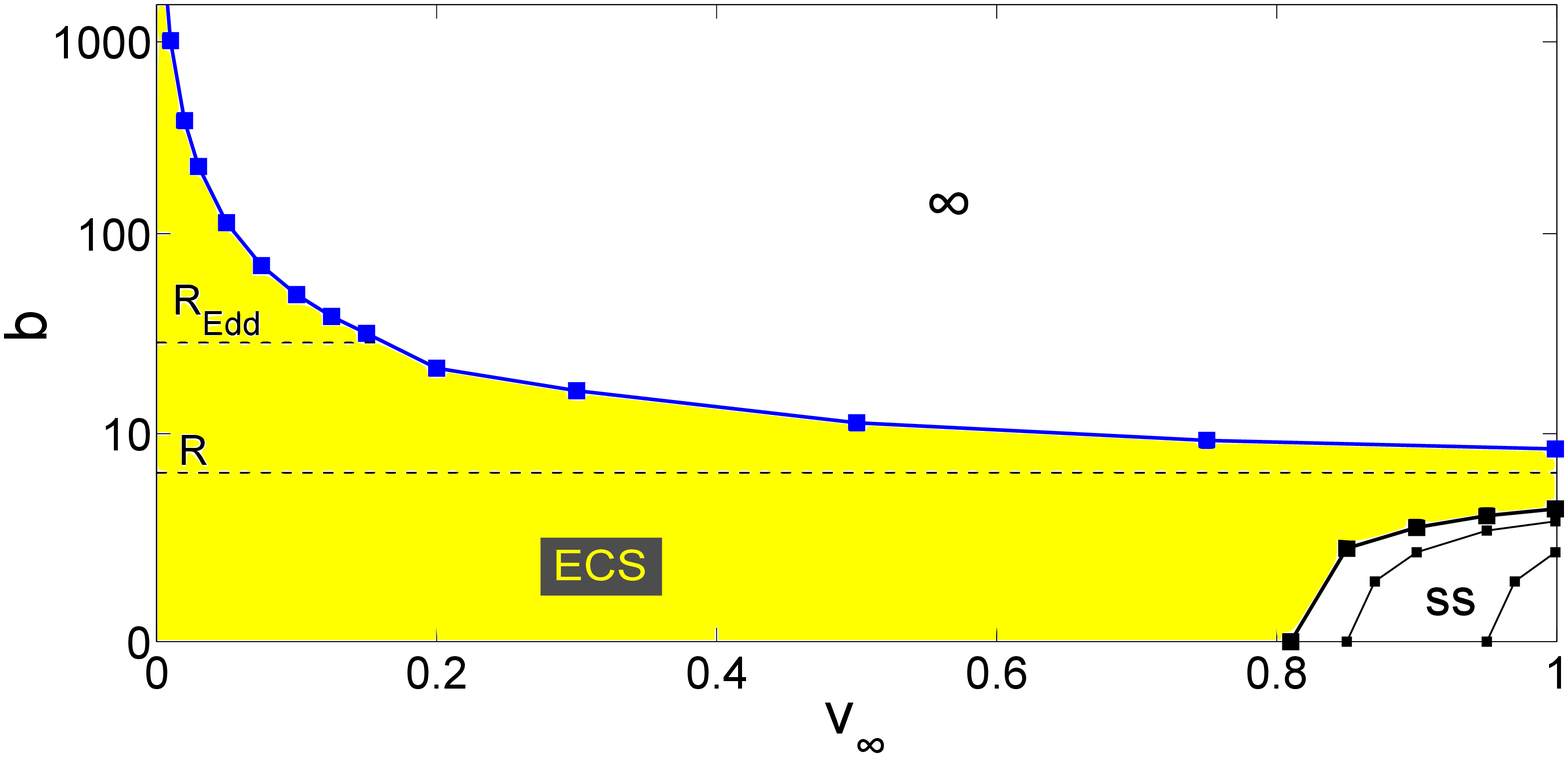}
\includegraphics[width=0.48\textwidth]{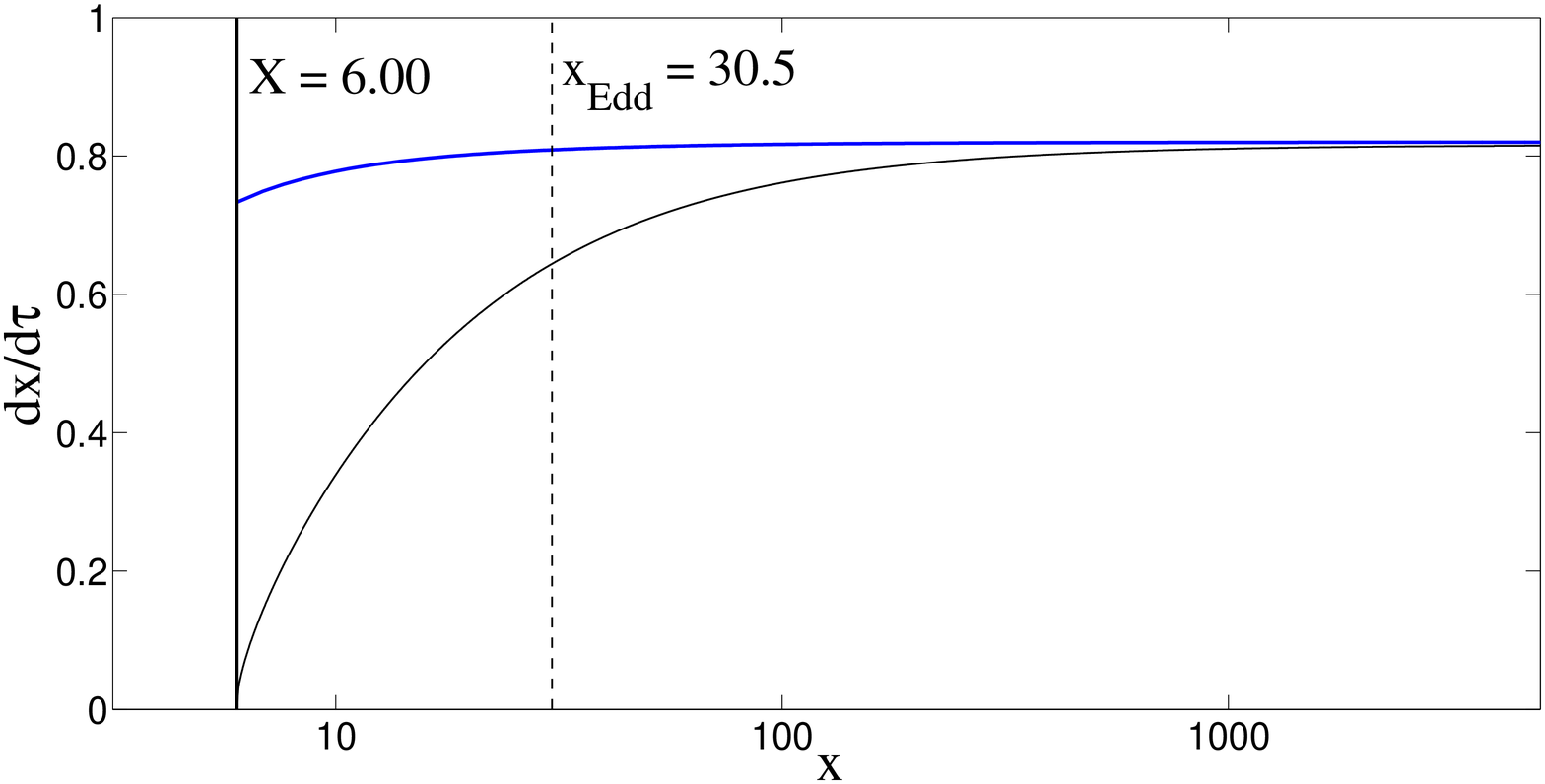}
\caption{The case of k=1.45. \emph{Top:} Same as Fig. 7.
\emph{Bottom:} The velocity as a function of radius for radial
infall ($b=0$) and $v_\infty=0.82c$, with and without radiation
 drag (bottom and top curves, respectively).}
\label{Fig:ss1}
\end{figure}

For a sufficiently low impact parameter, and a sufficiently high
incoming velocity at infinity, a test particle may penetrate
the Eddington sphere deeply enough to reach the stellar surface.
The minimum value of $v_\infty$ required to reach the star in the
case corresponding to Fig.~\ref{Fig:HL}, i.e., $k=1.49$ is shown as a
function of the impact parameter in the top panel of Fig.~\ref{Fig:ss}
(the monotonically growing function
depicted by the thick black line), for these velocities
the particles settle on the stellar surface with zero velocity.
This is a case of accretion not leading to any release of energy
on the stellar surface. If the velocity at infinity
is higher, the test particle will deposit some kinetic energy
at impact on the stellar surface. However, the deposited kinetic
energy will be a tiny fraction of the kinetic energy at infinity,
as the impact velocities are much lower than $v_\infty$
(thin black lines in Fig.~\ref{Fig:ss}).
This outcome is just the opposite of what occurs when radiation drag
and pressure may be neglected.

Other cases ($k=1.45$, and $k=1.25$) are illustrated in the bottom
panel of Fig.~\ref{Fig:ss} and in Fig.~\ref{Fig:ss1}.
The bottom panel of the latter figure illustrates the importance
of radial radiation drag. The top curve illustrates the changing
velocity of a test particle  accelerated by gravity and decelerated
by the radiation pressure 
[i.e., the term proportional to the $(t)(r)$ component of the stress tensor],
but not suffering from radiation drag. The bottom curve shows the
velocity computed with our full equations---clearly the resistance
offered by the radiation field slows down the particle very effectively.
Incidentally, the top curve illustrates another relativistic effect.
For a rapidly moving particle
the radiation pressure is enhanced by the ratio of $E/(mc^2)$
and it overwhelms the gravitational pull, so that the test particle
is decelerated well outside the Eddington sphere.
%
\section{Conclusions and discussion}
We have shown that a luminous star in Schwarzschild metric can
capture test particles from a wide class of orbits onto a spherical
surface of radius larger than the star itself.  The particles come
to rest on this sphere, because radiation drag eventually removes
all of their angular momentum, as shown by Oh, Kim, \& Lee (2011).
 The particles remain suspended above
the surface of the star, because every point on the capture sphere
corresponds to a position of equilibrium, the equilibrium
being stable in the radial direction (Abramowicz, Ellis, Lanza 1990)
and neutral in directions tangent to the sphere. The radius of this
``Eddington Capture Sphere" depends on the ratio of the stellar
luminosity to the Eddington luminosity (Eq. [1]).  Thus, accreting
compact stars, as well as stars which eject matter, may be surrounded
by a shell of matter at rest, as long as the luminosity of the star is
close to the Eddington luminosity (super-Eddington on the stellar
surface).

The trajectories of particles falling from infinity in the gravitational
field of a mildly super-Eddington compact star were investigated
in Section 6. We showed that radiation drag strongly reduces both
the azimuthal and the radial component of test particle velocities.
This leads to a moderate enhancement of the capture cross-section
by the Eddington sphere, relative to the cross-section for accretion
by a non-radiating star, and a drastic suppression of the cross-section
for actual accretion onto the stellar surface. Radiation drag also leads
to a dramatic decrease in the kinetic energy deposited at the surface.

It is interesting to speculate that the presence of
the Eddington Capture Sphere, an effect of Einstein's general
relativity, may play a role in the formation and ejection of shells of
matter by massive stars in the final stages of their evolution, such
as Wolf-Rayet stars, or the luminous blue variables (LBV stars).  In
principle, the capture sphere can exist at any radius outside the
star, if the stellar luminosity has the appropriate value. However,
the larger the radius of the sphere (in units of the Schwarzschild
radius), the more finely tuned the luminosity has to be---for a
$10R_G$ (5~Schwarzschild radii) capture sphere the luminosity at
infinity has to be $0.95 L_{\rm Edd}$, while for a 500 000 $R_G$ sphere
the luminosity has to be equal to the Eddington value to an accuracy
better than one part in a million.  While our discussion was couched
in terms of the Thomson scattering cross-section, similar results will
hold if other scattering or absorption processes contribute to
momentum transfer from the radiation field to the particles. The
capture sphere is located at that radius at which radiation pressure
on matter at rest balances the gravitational pull of the star. Outside
the sphere, the radiation pressure is too weak to balance gravity,
inside the sphere it overcomes the gravitational pull. 

It is expected that a more likely application of the Eddington capture
sphere will be found in the interpretation of the behaviour of
accreting neutron stars in low mass X-ray binaries. In particular, in
the Z sources, the inner radius inferred from both the kHz QPOs 
and the accretion-disk spectral
component have been determined to vary rapidly at a luminosity which is
thought to be close to the Eddington value (\cite{Yu,Lin}). 
In particular, the kHz QPOs vary with the normal branch oscillation
(NBO) phase, which may indicate
radiation induced rapid drifts of the innermost accretion flow.
 Perhaps these findings could be understood in terms of the expected
behaviour of accreting matter accumulating on (a sector of) the
Eddington sphere.  In the non-spherical geometry of accretion disks,
radiation scattered in non-radial directions can escape from the
system. Therefore the accreting fluid can ``levitate" above the star
only as long as it is optically thin. As more and more matter
accumulates at the capture radius of Eq. (1), the shell eventually
becomes optically thick and the outer layers of the fluid outweigh the
radiation pressure support, leading to rapid accretion of matter and
to the consequent evacuation of the region near the Eddington capture
sphere, allowing the process of accumulation to begin again.  One
would also expect the accreting fluid to be spread out to higher
altitudes (lower and higher values of the polar angle $\theta$) than
those subtended by the inner parts of the optically thick disk, as the
incoming fluid forms a ``puddle" on the surface of the Eddington
capture sphere.  A more detailed understanding of actual behaviour of
Z sources would require relaxing the assumption of spherical symmetry,
on which all the calculations in this paper were based.
Comparison of the theoretical
expectations with the observed NBO and its related
flaring branch oscillation (FBO) would be very interesting.

In the atoll sources, which are less luminous, the Eddington luminosity
is attained
during X-ray bursts. It would be worthwile to investigate whether 
the Eddington capture sphere plays a role in the evolution of
``radius expansion bursts" (e.g., \cite{Damen}).
In addition, in the
persistent emission at rather high luminosity levels $\sim 0.4\, L_{\rm Edd}$, 
a $\sim 7$ Hz QPO similar to the NBO in the Z sources was also observed in
the atoll source 4U 1820-30 (\cite{Wijnands}).
 The effect we are considering above is global, but may apply to
local properties of the accretion flow as well.

\section{Acknowledgements}

We thank Saul Rappaport for discussion and for providing
test orbits against which our codes were checked.
Research supported in part by Polish NCN grants NN203381436 and
UMO-2011/01/B/ST9/05439, as well as by a Swedish VR grant.
Research in the Instute of Physics at the Silesian University
in Opava supported by the Czech grant MSM 4781305903.
WK acknowledges the support of the National Science Foundation
and the hospitality of the Aspen Center in Physics, where a part
of this work was completed.

\end{document}